# Physics is the New Data

Sergei V. Kalinin,[1] Maxim Ziatdinov,[2,3] Bobby G. Sumpter,[3] and Andrew D. White[4]

[1] *Department of Materials Science and Engineering, University of Tennessee Knoxville, Knoxville, Tennessee 37996, USA*

[2] *Computational Sciences and Engineering Division, Oak Ridge National Laboratory, Oak Ridge, Tennessee 37831, USA*

[3] *Center for Nanophase Materials Sciences, Oak Ridge National Laboratory, Oak Ridge, Tennessee 37831, USA*

[4] *Department of Chemical Engineering, University of Rochester, Rochester, New York 14534, USA*

**The rapid development of machine learning (ML) methods has fundamentally affected numerous applications ranging from computer vision, biology, and medicine to accounting and text analytics.[1-3] Until now, it was the availability of large and often labeled data sets that enabled significant breakthroughs. However, the adoption of these methods in classical physical disciplines has been relatively slow, a tendency that can be traced to the intrinsic differences between correlative approaches of purely data-based ML and the causal hypothesis-driven nature of physical sciences. Furthermore, anomalous behaviors of classical ML necessitate addressing issues such as explainability and fairness of ML. We also note the sequence in which deep learning became mainstream in different scientific disciplines – starting from medicine and biology and then towards theoretical chemistry, and only after that, physics – is rooted in the progressively more complex level of descriptors, constraints, and causal structures available for incorporation in ML architectures. Here we put forth that over the next decade, physics will become a new data, and this will continue the transition from dot-coms and scientific computing concepts of 90ies to big data of 2000-2010 to deep learning of 2010-2020 to physics-enabled scientific ML.**

Neural networks and machine learning (ML) have been known since the early work on perceptrons in the 1950s. The introduction of the backpropagation algorithm in the 1980s laid the foundation for modern ML. However, the computational capabilities and lack of large labeled data sets limited the community to relatively shallow networks in applications such as molecular dynamics simulations, theory-experiment matching, and experimental data analysis.[4] The situation started to radically change after the 2000, when the availability of high-performance computing capabilities enabled large computational models and early experimental applications of neural networks were demonstrated. Similarly, the growth in internet technologies, including search engines and social networks, created a favorable environment for collecting and exploring large data sets, thus starting the deep learning revolution.

The striking aspect of deep learning (DL) is that, from a certain perspective, the field is just around ten years old. Indeed, the development of ImageNet by Fei-Fei Li et al. in 2009[5] has demonstrated the utility of large-annotated data sets, setting the foundation of the big data revolution. The ground-breaking paper by Krizhevsky in 2012,[6] introducing the potential of deep neural networks, then became the harbinger of the deep learning revolution of the last decade. Generative Adversarial Networks (GANs)[7] and Variational Autoencoders (VAEs)[8] appeared in 2014 (and by now, these preprints have ~80k citations). The concepts such as attention layers and transformers appeared in 2017.[9] Significant reinforcement learning-based advances in robotics and chip design are happening now.[10]

Throughout this past decade, much of this growth was fueled by big tech companies like Google, Facebook, Baidu, Amazon, Tencent, Uber, and Microsoft - and that in turn was predicated on their access to data and the necessity to derive actionable or monetizable insights from it. In many cases, the broad propagation of specific social media applications combined with the cell phone and mobile electronics ubiquity allows for centralized access to a significant fraction of all data in certain domains. However, equally important is that these advances started to strongly highlight the problems with data-based supervised learning methods. There are many aspects of these, ranging from adversarial attacks to the issues related to fairness and explainability - but they are ultimately linked to the fact that pure data-centric approaches are inherently limited in capturing the complexity of the real world. There are also limits to how far these can be scaled - simply as a matter of energy consumption, if nothing else.

These limits have been noticed by the ML world. Causal ML, metric learning, and, more generally, methods aiming at avoiding out of distribution effects are growing in popularity. The same is true for physics-informed machine learning from Hamiltonian to equivariant neural networks. Similarly, there is a rapid growth of interest in the ML studies of dynamic data generation processes such as autonomous microscopy or physical characterization instead of the analysis of the static benchmark data sets. This is a very recent trend - led by the Flatiron institute, Google Accelerated Science, the Vector Institute, and similar organizations that typically combine strong connections to both the industry and academic world.

A brief overview of the last ~3 years (at most) suggests that these developments can be a part of a more general trend – namely, ML methods progressively start to operate with concepts more abstract than just data sets. In other words, the ML field has evolved from "shallow" data-driven methods (think cats on the internet or hand-written digits) to methods utilizing more complex constructs such as graph and symbolic representations, invariances, positional embeddings – i.e., much deeper levels of scientific abstraction than simply data. This trend effectively brings in new concepts and connections to fundamental physics. Which brings us to a central concept of this perspective - *if we know that big data is necessary but limited in potential, then what will be the new data*? It seems that the existing (if not yet broadly realized) trend is the merger between physics and data sciences. It is also very remarkable that despite the tremendous developments in DL in biological, social, and economic sciences and theory, the progress in experimental sciences, including condensed matter physics, materials science, and chemistry, was relatively slow to develop.

Indeed, much of the original insights and developments in ML came from biological areas – starting from Hebb and more recently to Sejnowski and many others.[11] This was due to scientific reasons, e.g., the perceived similarity between NN architectures and brain/vision. The second aspect is that neural networks (NNs) were adopted much earlier and more broadly in biology and medicine. While one should be careful with speculation as to why, some factors include that biological areas are generally less susceptible to analytical derivations than physics, necessitating the data focus. Furthermore, physical sciences generally operate with complex causal and hypothesis-driven paradigms. Some of the leading scientists in the field such as Judea Pearl, one of the fathers of causal inference methods,[12] was a professor of physics and worked with superconductors before moving into biostatistics. Interestingly, this community is still very heavily bio-based – and very few (so far) causal ML groups have moved into physics, even though physics is much more amenable to interpretation in terms of causal mechanisms than biology.

Currently, the biological and medical areas are clearly the ones where classical ML methods can and are making major impact. However, these fields generally operate with relatively poorly defined entities and laws, making rigorous definitions difficult. As such, DL methods can be employed on this data directly. Comparatively, physical sciences offer rigorously defined entities, and often defined relationships between them structured via conservation laws, invariances, and causal relationships. In this case, ad hoc application of ML models based on data only may lead to unphysical results if the relationships are violated during training and fitting. In other words, while inference can violate relevant physical laws as a part of discovery (much like the introduction of complex numbers), the final results obtained need to comport to realistic physical constraints.

Applications of machine learning in modeling and simulation provided some of the earliest impacts, with examples spanning prediction of structure/property relationships to molten salt phase diagrams to molecular energy transfer, to multi-dimensional potential landscape design to toxicity prediction alongside numerous process control/optimization and correlative studies for spectroscopy. However, computational methods application is facilitated by the lack of out of distribution drift, meaning that the data generation processes are similar. By the same token, the causal chains in theoretical models are well understood.

Chemistry is a unique field because nearly all data are united upon the idea of molecular structure. Fields as far apart as protein structure and porous materials have a common underlying feature: atoms and bonds. Physics-informed ML is typically viewed as about the underlying model but in chemistry, the representations themselves imply equivariances. For example, features of molecular graphs and labels of atomic partial charge imply permutation equivariance. Features of Cartesian coordinates and labels of atomic partial charges imply translation and rotation invariance. Just choosing how to represent data is already connected with the physics of problems in chemistry.

The most significant recent advances in chemistry and biology AI have been accomplished with equivariant neural networks.[13-15] These neural networks are built for specific symmetry groups, and this enables vast reductions in training data because the symmetry structure is explicit rather than learned. This depends on the underlying data being from a known symmetry group. As deep learning moves to less data-rich areas of physics, having symmetry group attributes can significantly reduce the training data by being a strong inductive bias.

One of the strongest counterarguments the inductive biases of symmetry has been pretraining and other semi-supervised methods. For example, large language models (LLMs) appear to learn the structure of language without any explicit construction of grammar. LLMs are pretrained on large corpi, like billions of bioinformatics sequences or millions of natural language documents. The data is supposed to impart some "understanding" to the LLMs about language and then the LLMs have been quite successful on downstream tasks like predicting sentiment in a Tweet or predicting secondary structure in a protein.[16, 17] It appeared that unlabeled, unstructured data can be a scalable approach to learning. However, research groups have recently shown that training an LLM music data, gave a similar performance on downstream tasks about English. This seems to counter the hypothesis that large corpi are responsible for the success of pretraining. Lipton et al.[18] even went further and showed that even pretraining on randomly generated nonsense gives good performance. They hypothesized that it is the structure of the pretraining algorithm and tasks, rather than the underlying large data, that are responsible for the success of LLMs. Semi-supervised methods are being adapted to graphs[19] and point clouds,[20] so it remains to be seen if a purely data-driven approaches with enormous corpi can be successful.

Another remarkable development over the last several years is the rapid growth of the symbolic regression methods. In many areas of science, the presence of closed-form symbolic representations is perceived as the understanding of underlying mechanisms. The most well-known examples of these are equations of motions that stem from conservation laws of classical mechanics. However, in many scientific areas, the semiquantitative laws that define chemical reactivity of organic molecules, relate crystallographic structures to atomic radii, or define the direction of electron transfer between elements have been known for almost a century. The power of machine learning methods such as SISSO,[21] SinDy[22] and PySR[23] to derive these symbolic expressions from observational data is now opening new opportunities for scientific discovery. Particularly of interest is the transition of these methods towards active learning, in which case the ML algorithm interacts with the data generation process to effectively conduct scientific experiments.

Finally, applications of DL in experimental physical sciences presents the ultimate challenge. While it is generally accepted that the observed structures and processes are governed by the causal physical laws, the observations and experiments necessarily yield biased representation of reality. In many cases, the property of the observed object and observing system cannot be decoupled, with examples ranging from microscopy to nanoindentation. In classical instrumental sciences, the calibration and quantification of the observations in terms of robust material-specific descriptors is at premium. Similarly, experiments almost inevitably contain observational biases and confounders, severely limiting interpretation.

Indeed, one of the major limitations of modern DL models is their correlative nature. During the training stage, a DL model minimizes its loss objective by absorbing all the spurious correlations[24] – including confounding factors and selection biases – found in the training dataset. Because these correlations are not related to actual causal mechanisms, they can easily change between training and application domains. This commonly leads to poor performance of pre-trained models on data produced by distributions different from the training distribution (known as an out-of-distribution or OOD effect). For example, replacing a scanner in the electron microscopy experiment can drastically reduce the recognition capabilities of the DL model on the same sample.

Incorporating causality into the DL frameworks is still in the early stages, including works on invariant risk minimization[25] and causal representation learning.[26] However, none of the suggested to-date approaches have demonstrated reliable performance on real-world OOD data.

An alternative solution is to replace a model pre-trained on a static dataset with a model interacting actively with a data generation process. For example, in the active learning approach, a surrogate model – such as the Gaussian process or Bayesian neural network – takes the (sparsely) measured data as input and produces the expected values and associated uncertainties of the physical property of interest in the unmeasured parts of parameter space, which can then be used to derive the next measurement point. However, the standard active learning frameworks do not readily allow for incorporating different data modalities or prior physical knowledge. The recently introduced hypothesis learning framework[27] allowed a co-navigation of hypothesis and experimental spaces by incorporating probabilistic models of possible system behaviors and reinforcement learning policies into the active learning setup. It is based on the idea that a correct model of a system's behavior leads to a faster decay of the posterior predictive uncertainty and allows quickly learning the right model together with the overall data distribution using a relatively small number of measurements. Hypothesis learning can be extended to the active learning of structure-property relationships in the multi-modal experiments, where one can quickly learn the information channel with the best predictive capacity for the property of interest.

However, being able to pose causative, interventional, and counterfactual questions as well as integration of abstract concepts, should probably start at the interface between physical and computational sciences where arguably the data would have higher veracity.

**Acknowledgements:** The overall concept is based upon work supported by the U.S. Department of Energy (DOE), Office of Science, Basic Energy Sciences (BES), Materials Sciences and Engineering Division and was performed at the Center for Nanophase Materials Sciences, a US Department of Energy office of Science User Facility at Oak Ridge National Laboratory. Research reported in this work was supported by the National Institute of General Medical Sciences of the National Institutes of Health under award number R35GM137966 (A.D.W.). The authors gratefully acknowledge Kyle S. Cranmer for valuable discussions.